\documentclass{cjaa}                   

\usepackage{graphicx}                   
\input{epsf.sty}                        
\input{psfig.sty}                       

\usepackage{amsfonts}
\usepackage{amsmath}

 \font\sevenrm=cmr7 scaled 1000

\def\gsim{\;\lower4pt\hbox{${\buildrel\displaystyle >\over\sim}$}\;}
\def\lsim{\;\lower4pt\hbox{${\buildrel\displaystyle <\over\sim}$}\;}
\def\grls{\;\lower4pt\hbox{${\buildrel\displaystyle >\over <}$}\;}

\begin{document}

   \title{Outflows and inflows in astrophysical systems
}

   \volnopage{Vol.0 (200x) No.0, 000--000}      
   \setcounter{page}{1}          

   \author{Yue Shen
      \inst{1}\mailto{}
   \and Yu-Qing Lou
      \inst{1,2,3}
      }
   \offprints{Y. Shen}                   

   \institute{Physics Department, The Tsinghua Center for
Astrophysics (THCA), Tsinghua University, Beijing 100084, China\\
             \email{shenyue98@tsinghua.org.cn}
        \and
             National Astronomical Observatories, Chinese Academy
of Sciences, A20, Datun Road, Beijing 100012, China\\
        \and
             Department of Astronomy and Astrophysics, The University of
Chicago, 5640 S. Ellis Ave., Chicago, IL 60637 USA\\
          }

   \date{Received~~2004 month day; accepted~~2004~~month day}

   \abstract{
We seek for self-similar solutions describing the time-dependent
evolution of self-gravity systems with either spherical symmetry
or axisymmetric disk geometry. By assuming self-similar variable
$x\equiv r/at$ where $a$ is isothermal sound speed we find
self-similar solutions extending from the initial instant $t=0$ to
the final stage $t\rightarrow \infty$ using standard
semi-analytical methods. Different types of solutions are
constructed, which describe overall expansion or collapse,
envelope expansion with core collapse (EECC), the formation of
central rotationally supported quasi-equilibrium disk as well as
shocks. Though infinitely many, these self-similarity solutions
have similar asymptotic behaviors which may impose diagnosis on
the velocity and density structures in astrophysical systems.
   }

   \authorrunning{Y. Shen \& Y.-Q. Lou }            
   \titlerunning{Self-similar solutions}  

   \maketitle

%
%
\section{Introduction}           
\label{sect:intro}
Far away from initial and boundary conditions, dynamical evolution
of fluid systems may lead to self-similar phases with physical
variable profiles shape-invariant and magnitudes properly scaled
(Sedov 1959; Landau \& Lifshitz 1959). Similarity methods, which
transform partial differential equations (PDEs) to ordinary
differential equations (ODEs), greatly simplify nonlinear
problems. For spherical systems, several well-known self-similar
solutions were found to describe the collapse of an isothermal
cloud in the context of star formation (Larson 1969; Penston 1969;
Shu 1977; Hunter 1977; Whitworth \& Summers 1985). The first is
the LP solution independently found by Larson (1969) and Pensten
(1969); the second is the ``expansion wave collapse'' solution
(EWCS) found by Shu (1977), who also discovered other solutions
with central point mass from the initial instant $t=0$ to the
final stage $t\rightarrow +\infty$; Hunter (1977) discovered
infinitely many discrete analytic solutions within the
pre-catastrophic period which share with the LP solution that they
are regular at $x\rightarrow 0^{-}$ although some of these
solutions can not be continued in the post-catastrophic period
from $t=0$ to $t\rightarrow +\infty$. Hunter's results were
extended by Whitworth \& Summers (1985, hereafter WS), who allowed
weak discontinuities across the sonic critical line and hence
constructed continuous bands instead of Hunter's discrete
solutions. But their solutions with weak discontinuities were
criticized by Hunter (1986) that are only of mathematical interest
and are physically unstable. Recently, Lou \& Shen (2004,
hereafter LS04) re-examined this classic problem and derived new
solutions in the `semi-complete space' ($0<t<+\infty$), in
contrast to the `complete space' ($-\infty<t<+\infty$) taken by
Hunter (1977) and WS.

The realistic situation may deviate from the spherical collapse.
In essence, the presence of rotation or magnetic fields will cause
the collapsing system to be more and more flattened (e.g.,
Nakamura, Hanawa, \& Nakano 1995, hereafter NHN) so that the
system will later evolve disk-like. Models considering
self-similar evolutions in a thin disk have been studied in many
papers (e.g., NHN; Li \& Shu 1997; Saigo \& Hanawa 1998;
Krasnopolsky \& K\"{o}nigl 2002) with or with out rotation and
magnetic fields. Axisymmetry is assumed
in the above papers to simplify the problem. 

In this work, we plan to establish the whole solution structure
for an isothermal fluid system with either spherical symmetry or
axisymmetric disk geometry within the period from $t=0$ to
$t\rightarrow\infty$. In comparison with the previous solutions
found by other authors, we particularly emphasize the existence of
such solutions that the innermost part (core) is collapsing
towards free-fall while simultaneously the outer part (envelope)
is expanding and approaching constant wind. However, the overall
collapse region is expanding in a self-similar manner. These
envelope expansion with core collapse (EECC) solutions exist in
the spherical and non-rotating disk cases but not the rotating
disk case (see Section 3.2). Furthermore, as an non-trivial
extension we can incorporate purely azimuthal magnetic fields in
the disk cases (details are not discussed here).


\section{Spherical Cases}
\label{sect:Obs}
The basic formalism for the problem with spherical symmetry is
well-established by many authors. By the definition of
self-similar variable $x\equiv r/at$ where $r$, $a$, $t$ are the
radial coordinate, isothermal sound speed and time coordinate
respectively, the reduced ordinary differential equations (ODEs)
are as follows
\begin{equation}
\begin{split}
&[(x-v)^2-1]\frac{dv}{dx}
=\bigg[\alpha(x-v)-\frac{2}{x}\bigg](x-v)\ ,\
[(x-v)^2-1]\frac{1}{\alpha}\frac{d\alpha}{dx}
=\bigg[\alpha-\frac{2}{x}(x-v)\bigg](x-v)\ ,\\
&m=x^2\alpha(x-v)\ ,
\end{split}
\end{equation}
where $\alpha(x)$, $v(x)$, $m(x)$ are the reduced density, radial
bulk velocity and enclosed mass, respectively. The physical
density $\rho$, radial bulk velocity $u$ and enclosed mass $M$ are
obtained by the similarity transformation
\begin{equation}
\begin{split}
\rho(r,t)=\frac{\alpha(x)}{4\pi Gt^2},\ \ \ \ u(r,t)=av(x),\ \ \ \
M(r,t)=\frac{a^3t}{G}m(x).
\end{split}
\end{equation}

By expanding the solutions in power serials at large and small
$x$; and with careful treatment of the sonic critical line one may
construct various types of self-similar solutions which are
present in LS04. In addition to the shock-free solutions present
in LS04, we can also construct shocked similarity solutions, which
are applied to the star-forming regions and H {\sevenrm II}
regions (Shen \& Lou 2004).

Among various solutions, we are particularly interested in such
solutions, for which the innermost part (core) is collapsing and
approaching free-fall towards the center while simultaneously the
outer part (envelope) is expanding and approaching a constant
wind. The core-mass accretion rate is constant. The shock which
connects the inner part and the outer part is propagating outward
at a constant speed.

\section{Axisymmetric Disk cases}
\label{sect:data}
For axisymmetric disk cases we introduce similarity
transformation:
\begin{equation}
\begin{split}
&x=\frac{r}{at}\ ,\ \Sigma(r,\ t)=\frac{a\alpha(x)}{2\pi Gt}\ ,\
u(r,\
t)=av(x)\ ,\ C_A=aq(x)\ ,\\
&M(r,\ t)=\frac{a^3t}{G}m(x)\ ,\ j(r,\ t)=\beta a^2tm(x)\ ,\ \Phi(r,\ t)=a^2\phi(x)\ ,\\ 
\end{split}
\end{equation}
where
\begin{equation}\label{beqn1}
\begin{split}
\phi(x)=-\frac{1}{2\pi x}\int_0^{\infty}{\cal
X}(\xi/x)\alpha(\xi)\xi d\xi\ ,\ {\cal
X}(X)=\oint\frac{d\psi}{(1+X^2-2X\cos\psi)^{1/2}}\ ,
\end{split}
\end{equation}
and $\alpha$, $v$, $q$, $m$, $\phi$ are reduced quantities which
are dimensionless functions of similarity independent-variable $x$
and $\Sigma$, $u$, $C_A$, $M$, $j$, $\Phi$ are surface mass
density, radial bulk velocity, Alfv\'{e}nic speed, enclosed mass,
specific angular momentum and gravitational potential
respectively. The specific angular momentum is proportional to the
enclosed mass because their ratio is assumed to be spatially
uniform at the initial instant $t=0$ and conserved during the
successive evolution (Li \& Shu 1997; Saigo \& Hanawa 1998). Using
these reduced variables we derive the ordinary differential
equations (ODEs) as
\begin{equation}\label{beqn2}
[(x-v)^2-1-q^2]\frac{dv}{dx}=\bigg(f-\frac{\beta^2
m^2}{x^3}-\frac{1}{x}\bigg)(x-v)+\frac{q^2}{2}\bigg(1-\frac{2v}{x}\bigg)\
,
\end{equation}
\begin{equation}\label{beqn3}
[(x-v)^2-1-q^2]\frac{1}{\alpha}\frac{d\alpha}{dx}=f-\frac{\beta^2
m^2}{x^3}-\frac{(x-v)^2}{x}+\frac{q^2}{2x}\frac{3x-4v}{x-v}\ ,
\end{equation}
\begin{equation}\label{beqn4}
[(x-v)^2-1-q^2]\frac{1}{q}\frac{dq}{dx}=\frac{f}{2}-\frac{\beta^2m^2}{2x^3}+\frac{q^2-2}{4(x-v)}
+\frac{v}{2x}\frac{2-(x-v)^2}{x-v}\ ,
\end{equation}
where $f\equiv d\phi/dx$ together with the following auxiliary
equations:
\begin{equation}\label{beqn5}
m=x\alpha(x-v)\ ,
\end{equation}
and equation (4). Equations (\ref{beqn1})-(\ref{beqn5}) are the
full set of ODEs to be solved. In particular, equation
(\ref{beqn2})-(\ref{beqn4}) are simultaneously solved for reduced
density, reduced radial velocity and Alfv\'{e}nic Mach number. In
the monopole approximation, the reduced gravitational force is
approximated as $f\sim m/x^2$.

There exists one exact solution for the ODE set, which is the
rotational equilibrium state of a (magnetized) singular isothermal
disk (MSID):
\begin{equation}
\begin{split}
&v=0\ , \ \alpha=\frac{1+D^2-Q^2/2}{x}\ ,\
f=\frac{1+D^2-Q^2/2}{x}\
,\ m=(1+D^2-Q^2/2)x\ ,\\
&\beta=\frac{D}{1+D^2-Q^2/2}\ ,\ q\equiv\hbox{constant}=Q\ .
\end{split}
\end{equation}

Without magnetic fields ($q=0$) and in the monopole approximation,
we reduce the basic equation set (\ref{beqn1})-(\ref{beqn5}) into
\begin{equation}\label{ode2a}
\begin{split}
&[(x-v)^2-1]\frac{dv}{dx}=\frac{\alpha
(x-v)[1-\beta^2\alpha(x-v)]-1}{x}(x-v)\ ,\\
&[(x-v)^2-1]\frac{1}{\alpha}\frac{d\alpha}{dx}=\frac{\alpha[1-\beta^2\alpha(x-v)]-(x-v)}{x}(x-v)\
,\ m=x\alpha(x-v)\ ,
\end{split}
\end{equation}
with the asymptotic solution when $x\rightarrow +\infty$ as
\begin{equation}\label{asyinf1}
\begin{split}
&v\rightarrow
V+\frac{1-A+\beta^2A^2}{x}+\frac{V(1-\beta^2A^2)}{2x^2}
+\frac{2V^2+(A-1-\beta^2A^2)(A-4)}{6x^3}\ ,\\
&\alpha\rightarrow\frac{A}{x}+\frac{A(1-A+\beta^2A^2)}{2x^3}\ ,\
m\rightarrow Ax\ .
\end{split}
\end{equation}
For non-rotating disks we simply set $\beta=0$ in the above
equations.

\subsection{Non-rotating Disks Without Magnetic Fields}
For non-rotating $\beta=0$ cases, we have asymptotic solutions
when $x\rightarrow 0$ as either
\begin{equation}\label{asyx01}
v\rightarrow -\bigg(\frac{2m_0}{x}\bigg)^{1/2}\ ,\
\alpha\rightarrow \bigg(\frac{m_0}{2x}\bigg)^{1/2}\ ,\
m\rightarrow m_0\ ,
\end{equation}
which stands for free-fall, or else
\begin{equation}\label{asyx02}
v\rightarrow \frac{x}{2}\ ,\ \alpha\rightarrow B\ ,\ m\rightarrow
Bx^2\ ,
\end{equation}
which stands for core-expansion. And the expansions of solutions
near the sonic critical line $x-v=1$ are
\begin{equation}\label{eigensolution}
\begin{split}
v=x_*-1+C_{\mp}(x-x_*)+...\ ,\qquad
\alpha=1+(C_{\mp}-1/x_*)(x-x_*)+...\ ,
\end{split}
\end{equation}
where
\begin{equation}
C_{\mp}=\frac{1\mp\sqrt{1+2(1/x_*^2-1/x_*)}}{2}\ ,
\end{equation}
with -minus and -plus signs denoting type 1 and type 2 solutions
in reference to the spherical case (Hunter 1977; WS; LS04). Figure
1 shows several example solutions for the non-rotating case.

   \begin{figure}
   \plottwo{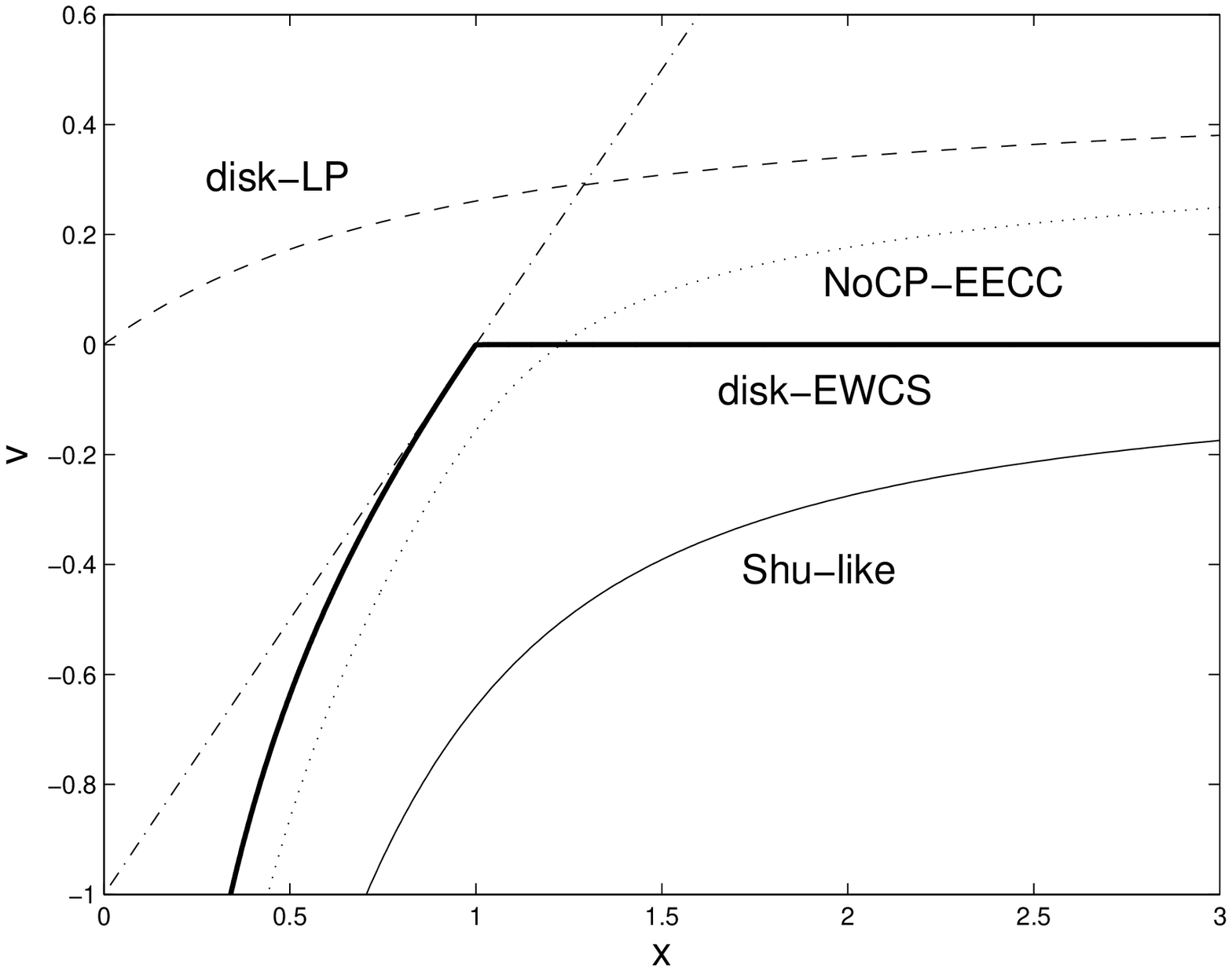} {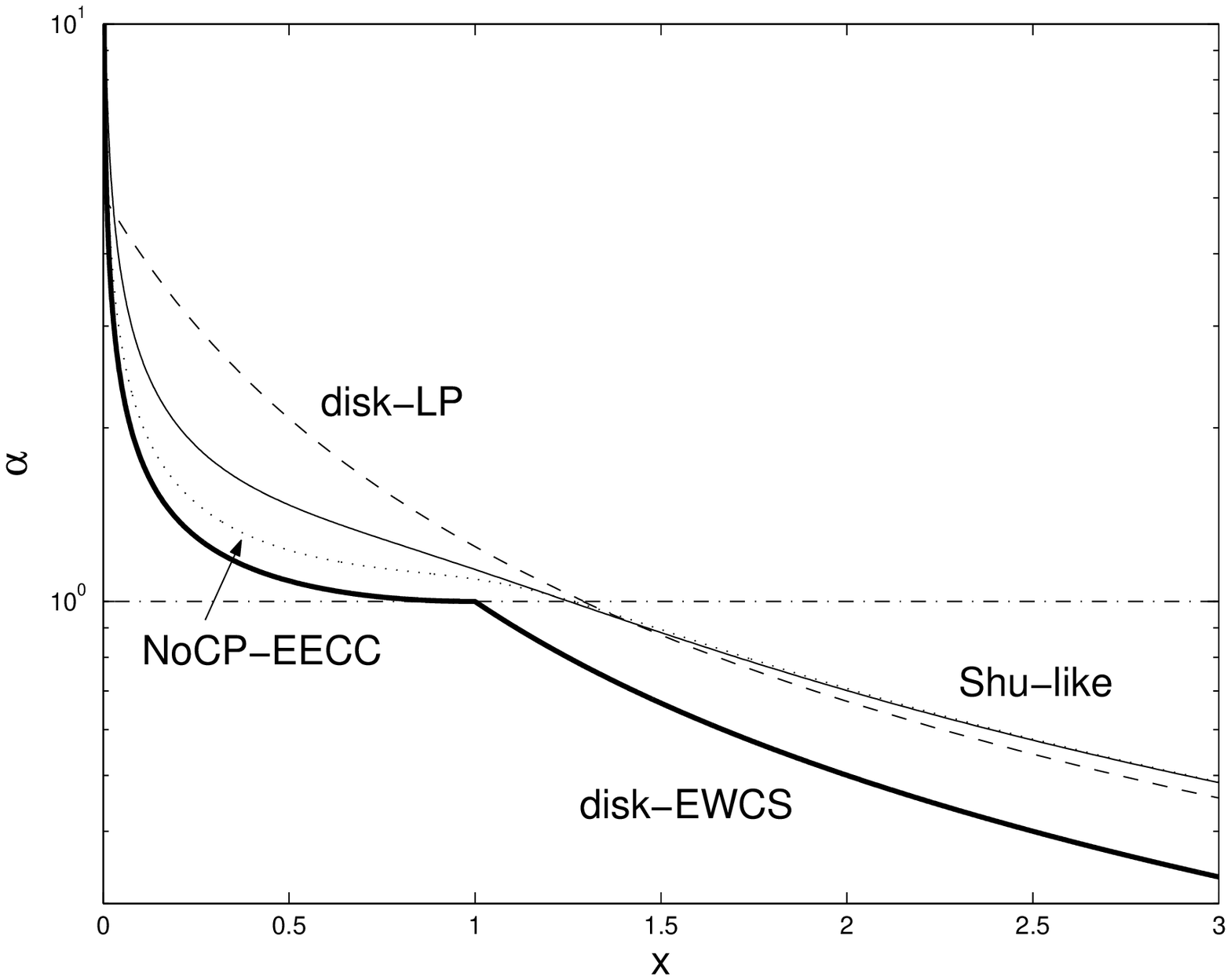}
   \caption{Example solutions in non-rotating disks without magnetic fields.
   Left: Example solutions of $v(x)$ versus similarity
variable $x$. The dash-dotted line is the sonic critical line
$x-v=1$. The light solid line is one of the Shu-like solutions and
the heavy solid line is the disk-EWCS. The dotted line is one of
the disk-EECC solutions which do not cross the sonic critical
line. The dashed line is the disk-LP solution. Right: Example
solutions of $\alpha(x)$ versus similarity variable $x$. Notations
and line-types are the same as left.}
   \label{Fig:plot2}
\end{figure}

\subsection{Rotating Disks Without Magnetic Fields}
For rotating $\beta>0$ cases, notice from equation (5) that the
term of centrifugal force diverges as $x^{-3}$ if $m\rightarrow
m_0$, faster than other terms. So for rotating disks the material
will not fall freely to the center as indicated by (\ref{asyx01}).
Instead, we have asymptotic solutions when $x\rightarrow 0$ as
either
\begin{equation}\label{asym0regular}
\begin{split}
&v(x)\rightarrow \frac{x}{2}-\frac{B}{12}x^2+\frac{1}{32}
\bigg[1+\bigg(\frac{1}{3}+\beta^2\bigg)B^2\bigg]x^3+...\ ,\\
&\alpha(x)\rightarrow
B-\frac{B^2}{2}x+\frac{B}{8}\bigg[1+\bigg(\frac{5}{3}+\beta^2\bigg)B^2\bigg]x^2+...\
;
\end{split}
\end{equation}
or the quasi-equilibrium asymptotic solution
\begin{equation}\label{asym0static}
v(x)\rightarrow \frac{D^2-1}{(1+D^2)(1+D^2+\lambda)}\eta
x^{\lambda+2}\ ,\  \alpha(x)\rightarrow \frac{1+D^2}{x}+\eta
x^{\lambda}\ ,\  m(x)\rightarrow (1+D^2)x\ ,
\end{equation}
where
\begin{equation}
\lambda=\frac{-3+\sqrt{4D^2-3}}{2}>-1\ ,\qquad
D=\frac{1+\sqrt{1-4\beta^2}}{2\beta}>1\ ,
\end{equation}
and $\eta$ is a constant parameter. Asymptotic solution
(\ref{asym0static}) is actually the quasi-equilibrium state with
the leading order as the exact equilibrium SID solution and is
rotating supersonically with rotational Mach number $D$ exceeds
unity.

By the isothermal shock jump condition (Tsai \& Hsu 1995; Saigo \&
Hanawa 1998; Shu et al. 2002)
\begin{equation}
(v_d-x_s)(v_u-x_s)=1\ ,\qquad
\frac{\alpha_u}{\alpha_d}=(v_d-x_s)^2\ ,
\end{equation}
we can connect the inner quasi-equilibrium solution to the outer
envelope which approaches constant winds or inflows or breezes. In
the above equations $x_s$ is the shock location fixed in the
similarity coordinates which suggests that the shock moves outward
at constant speed $x_sa$, i.e., the physical radius of the shock
discontinuity increases linearly with time.

\section{Discussion}
The major difference of our self-similar solutions from those by
other authors is that our solutions include not only those with
constant inflow speeds or being quasi-static at large radii but
also those with constant winds at large radii. Spatially, the
novel solutions describe both collapsing part and expanding part,
connected with or without shocks. Temporally, the fluid element at
all radii is decelerated from expansion to collapse and the
stagnation points as well as shock fronts propagate outward at
constant speeds. The density and velocity profiles at large or
small radii have well-defined power-law behaviors on radius, which
thus can be used as useful input parameters in numerical radiation
transfer codes. Another important aspect of these self-similar
solutions is the core-mass accretion rate, which is constant. So
the mass of the central object (either a proto-star or a black
hole) increases linearly with time. We have extended the range of
the core-mass accretion rate, which is only of specific values for
several known self-similar solutions (i.e., for the EWCS, the
core-mass accretion rate $m_0=0.975$).

The astrophysical applications of these self-similar solutions may
appear in various circumstances: systems involving accretions and
outflows. To be more specific, the evolution of young stellar
objects (YSOs), H {\sevenrm II} regions around OB stars and
planetary nebulae (PNe) may experience certain stages which can be
described by these self-similar solutions (Shen \& Lou 2004). By
incorporating more realistic ingredients such as the polytropic
equation of state, the toroidal environment and even general
relativity, these solutions originally derived in isothermal
spheres can be generalized to describe more complicated cases
(e.g., core-collapse supernovae). In reference to our novel EECC
solutions, we may propose the following scenario: the envelope at
large radii is initially expanding due to external heating (i.e.,
from supernovae or evolved stars near star-forming regions); while
the core region begins to collapse due to gravitational
instabilities; the successive evolution of the core-envelope
system may evolve into a self-similar state described by one of
the EECC solutions. Such a situation may be applicable to the
protostellar evolution where core accretion and large-scale
outflows may concur. For planetary nebulae, it is interesting to
note that in the classic interacting stellar wind (ISW) model, a
much faster stellar wind from the core catches up with a slower
dense wind--the remnant of the asymptotic giant branch phase--and
together they form shocks; while our shocked EECC solutions imply
that while the shocked expanding envelope is creating the observed
planetary nebula, the core (i.e., the proto-white dwarf) is
continuously accreting material. If it can accrete enough material
to exceed the Chandrasekhar limit, there might be the possibility
of igniting a Type Ia supernova explosion, without the requirement
for a companion star!

\label{lastpage}

\end{document}